\documentclass[CEJP,DVI]{cej} % use DVI command to enable LaTeX driver
\usepackage{layout}
\usepackage{amsmath}
\usepackage{textcomp}
\usepackage{hyperref}

\title{{\tt VISHNU} hybrid model for the viscous QCD matter at RHIC and LHC energies}

\articletype{Editorial}

\author{Huichao Song\inst{1}\email{song.199@osu.edu},
}

\institute{
     \inst{1} Department of Physics, The Ohio State University,\\
     191 West Woodruff Avenue, Columbus, Ohio 43210, USA
      }

\abstract{In this proceeding, we briefly describe the viscous hydrodynamics + hadron cascade hybrid model {\tt VISHNU}
for relativistic heavy ion collisions and report the current status on extracting the QGP viscosity from elliptic flow data. }


\begin{document}
\maketitle

%% ###################################################################

\section{ {\tt VISHNU} hybrid model:}
Viscous hydrodynamics is a useful tool to describe the expansion of the QGP fireball created in relativistic heavy ion collisions. Previous viscous hydrodynamic simulations reveal that elliptic flow $v_2$ and triangular flow $v_3$ are very sensitive to the shear viscosity. Even the ``minimal" specific shear viscosity given by the KSS bound $\eta/s = 1/{4\pi}$~\cite{Policastro:2001yc}  can lead to a significant suppression of $v_2$~\cite{Song:2007fn,Romatschke:2007mq} and $v_3$~\cite{Schenke:2010rr,Qiu:2011hf}. In principle, this allows for an extraction of the QGP viscosity from viscous hydrodynamic calculations by tuning $\eta/s$ to fit the experimental $v_2$ and $v_3$ data at RHIC and LHC. In practice, this requires excellent control of other inputs in the theoretical calculations, which include the initial conditions, the equation of state (EOS),  hadronic dissipative effects, and a proper description of the chemical and thermal freeze-out in the hadronic stage~\cite{Song:2008hj}.

For a more realistic description of the late hadronic stage, we developed the hybrid code {\tt VISHNU}~\cite{Song:2010aq}, which couples viscous hydrodynamics for the QGP expansion with a hadron cascade model for the late hadronic evolution at a switching temperature $T_{sw}$. {\tt VISHNU} consists of three parts: {\tt VISH2+1} solves the Viscous Israel-Stewart Hydrodynamic equations in 2+1 space-time dimensions, with an assumption of longitudinal boost invariance~\cite{Song:2007fn}, {\tt H2O}~\cite{Song:2010aq} converts the hydrodynamic output into particles profiles for further hadron cascade ({\tt UrQMD}) propagation through a monte-carlo event generator based on the modified Cooper-Frye formula, and {\tt UrQMD} solves the Boltzmann equation for a variety of hadron species with flavor-dependent cross-sections~\cite{Bass:1998ca}.  By describing the hadronic rescattering and freeze-out procedure  microscopically, {\tt VISHNU} improves earlier, purely hydrodynamic models and eliminates the additional adjustable parameters required for the transport and freeze-out characteristics of the hadron phase, making it possible for a reliable extraction of the QGP viscosity from the experimental data.

The default switching temperature in {\tt VISHNU} is set to $T_{sw}= T_{ch}=165 \texttt{ MeV}$, which is close to the QGP phase transition temperature and the chemical freeze-out temperature at RHIC energies. Using a partial chemical equilibrium EOS as input, we investigated whether the microscopic Boltzmann approach can be emulated by the macroscopic hydrodynamic approach by varying the switching temperature $T_{sw}$ in {\tt VISHNU} simulation, and found a strong $T_{sw}$-dependence of the elliptic flow (with a constant $\eta/s$ as input). We then extracted a temperature dependent effective shear viscosity $(\eta/s)^{eff}(T)$ for {\tt UrQMD} and found that pure viscous hydrodynamics with $(\eta/s)^{eff}(T)$ as input could nicely fit the  $p_T$-spectra and differential elliptic flow $v_2(p_T)$ from {\tt VISHNU}. However, the extracted $(\eta/s)^{eff}(T)$ strongly depends on the pre-hydrodynamic history, particulary the chosen value of the QGP shear viscosity. It therefore does not represent the intrinsic transport properties of the hadronic matter, but a parameter that reflects some memories of the QGP transport properties. We conclude that
there exists no switching window below $T_{ch}$ in {\tt VISHNU}, where viscous hydrodynamics can replace the hadron cascade and remain predictive power; Both components of the hybrid model are needed for a quantitative and realistic description of the dynamical evolution of the QGP fireball created in relativistic heavy ion collisions~\cite{Song:2010aq}.

\section{The QGP shear viscosity at RHIC and LHC energies}

During the evolution of the QGP fireball, the shear viscosity controls the efficiency to convert the initial spacial eccentricity $\varepsilon_x$ to the fluid momentum anisotropies $\varepsilon_p$. However, the distribution of the final momentum anisotropies to the flow patterns $v_2(p_T)$ for different hadron species strongly depends on the chemical composition and the accumulated radial flow in the late hadronic stage. In Ref.~\cite{Song:2010mg,Song:2011hk}, we proposed to use the integrated elliptic flow $v_2^{ch}$ for all charged hadrons to extract the QGP shear viscosity since it is directly related to the fluid momentum anisotropies $\varepsilon_p$ and less sensitive to other details in {\tt VISHNU} calculation: such as the chosen bulk viscosity, the detailed form of the viscous $\delta f$ corrections, etc. Ref.~\cite{Song:2010mg} also shows that the theoretical $v_2^{ch}/\varepsilon_x$ curves as a function of multiplicity density per overlap area $\mathrm{dN_{ch}/(dy S)}$  are universal, which depend only on the specific QGP shear viscosity $(\eta/s)_{QGP}$, but not on the details of the initial conditions. Furthermore, pre-equilibrium flow and bulk viscosity only slightly affect these theoretical curves, which are at the order of 5\% or below 5\%. It thus preferable to extract the QGP viscosity from a comparison between the theoretical and experimental $v_2^{ch}/\varepsilon_x-\mathrm{dN_{ch}/(dy S)}$ curves, as shown in Fig. 1.  The colored lines with symbols  are the {\tt VISHNU} results with  different $(\eta/s)_{QGP}$ as input. Left and right panels correspond to  two different
smoothed initial conditions, obtained from averaging a large number of fluctuating initial entropy densities (given by MC-Glauber or MC-KLN models) by aligning the participant plane for each event. For the experimental curves, $v_2$ and $\mathrm{dN_{ch}/dy}$ are experiment measurements, while $\varepsilon$ and $S$ are theoretical inputs obtained from the MC-Glauber and MC-KLN models. This leads to the differences in magnitude and slight changes in  slope for the two experimental curves shown in Fig.1 left and right. As a result, the extracted value of $(\eta/s)_{QGP}$ from these two panels changes by a factor of 2 mainly due to the different $\varepsilon_x$ from MC-KLN and MC-Glauber. Recent event-by-event viscous hydrodynamic simulations showed that the triangular flow $v_3$ is also sensitive to the QGP shear viscosity~\cite{Schenke:2010rr,Qiu:2011hf} which suggests that, in the near future, the combined analysis of $v_2$ and $v_3$ from {\tt VISHNU} could yield an even more precise value of $(\eta/s)_{QGP}$ and may reduce the ambiguities in initial conditions from the hydrodynamic side. At this moment, we take the current uncertainties from the initial conditions and conclude from Fig.1 that $1 < 4\pi(\eta/s)_{QGP} < 2.5$ for the temperature region $T_c < T < 2T_c$ probed at RHIC.

%%%%%%%%%%%%%%%%%%%%%%%%%%%%%%%% Fig. 1 %%%%%%%%%%%%%%%%%%%%%%%%%%%%%%%%%%%%%%
\begin{figure}[t]
\begin{center}
 \includegraphics[width=0.8\linewidth,clip=,angle=0]{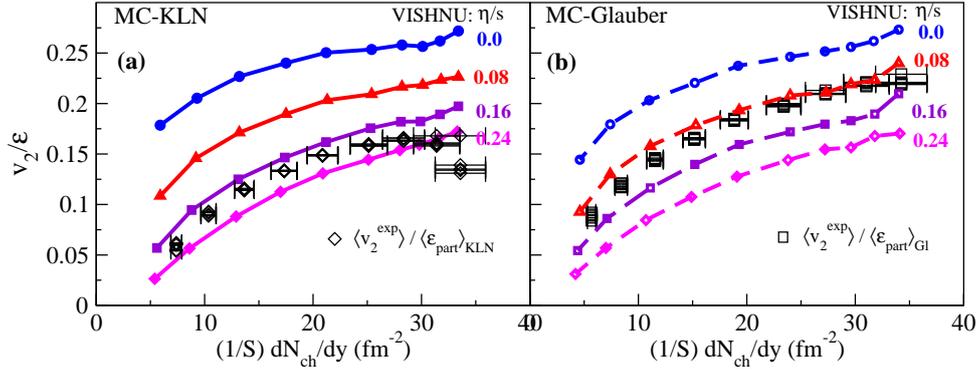}
\caption{\label{F1} (Color online) eccentricity-scaled elliptic flow as a function of final multiplicity per area\cite{Song:2010mg}.
}
\end{center}
\end{figure}
%%%%%%%%%%%%%%%%%%%%%%%%%%%%%%%%%%%%%%%%%%%%%%%%%%%%%%%%%%%%%%%%%%%%%%%%%%%%%%

With $(\eta/s)_{QGP} \simeq (1/4\pi)$ for MC-Glauber and $(\eta/s)_{QGP} \simeq (2/4\pi)$ for MC-KLN extracted from the
$p_T$ integrated $v_2$ for all charged hadrons, {\tt VISHNU} yields an excellent description of the $p_T$-spectra and differential elliptic flow $v_2(p_T)$ (for $p_T <2 GeV$)  for all charged hadrons and identified hadrons at different centrality bins measured in 200 A GeV Au+Au collisions at RHIC~\cite{Song:2011hk}, showing the power of the newly developed hybrid model and the robustness of the above extraction of the QGP shear viscosity.

The successful fit of the RHIC low $p_T$ data allows for constrained LHC predictions from {\tt VISHNU}. After extrapolating our calculation to the Pb+Pb collisions at LHC and comparing with recent flow data from the ALICE Collaboration, we found that
the $(\eta/s)_{QGP}$ extracted from RHIC slightly over-predicts the LHC data. After increasing  $(\eta/s)_{QGP}$ from $2/(4\pi)$ to $2.5/(4\pi)$ (for MC-KLN), {\tt VISHNU} gives a better description of the integrated and differential elliptic flow data for all charged hadrons at LHC~\cite{Song:2011qa}. A further simulation from {\tt VISHNU} also yields a reasonable description of the $v_2(p_T)$ for identified hadrons such as pion, kaon and protons at different centralities~\cite{Heinz:2011kt}. These results suggest that the \emph{averaged} specific QGP shear viscosity (or the assumed constant $(\eta/s)_{QGP}$) at LHC energies is slightly larger than at RHIC energies. However, a further study assuming different temperature dependencies for $(\eta/s)|_{QGP}$ shows that one cannot uniquely constrain the form of $(\eta/s)|_{QGP} (T)$ by fitting the spectra and $v_2$ alone. Based on our current understanding, the question on whether the QGP fluid is more viscous or more perfect in the temperature regime reached by LHC energies is still open~\cite{Song:2011qa}.\\

{\bf Acknowledgments:} The author thanks the collaborations and discussions from S. Bass, U. Heinz, T. Hirano and C. Shen for the work covered in this proceeding. This work was supported by the U.S.\ Department of Energy under grants
No. DE-AC02-05CH11231, \rm{DE-SC0004286} and (within the framework of the JET Collaboration) No. \rm{DE-SC0004104}. Extensive computing resources provided by the Ohio
Supercomputing Center are gratefully acknowledged.

\end{document}